\documentclass[preprint,prc,aps,showpacs,showkeys]{revtex4}
\usepackage{epsfig}
\begin{document}
\title{An Investigation of the Coupling Potential by means of S-matrix Inversion}
\author{I. Boztosun}
\address{Department of Physics, Erciyes University, Kayseri Turkey.}
\author{R.S. Mackintosh}
\address{Department of Physics, Open University, Milton Keynes UK.}
\date{\today}
\begin{abstract}
We investigate the inelastic coupling interaction by studying its
effect on the elastic scattering potential as determined by
inverting the elastic scattering $S$-matrix. We first address the
effect upon the real and imaginary elastic potentials of including
excited states of the target nucleus. We then investigate the
effect of a recently introduced novel coupling potential which has
been remarkably successful in reproducing the experimental data
for the $^{12}$C+$^{12}$C, $^{12}$C+$^{24}$Mg and
$^{16}$O+$^{28}$Si reactions over a wide range of energies.  This
coupling potential has the effect of deepening the real elastic
potential in the surface region, thereby explaining a common
feature of many phenomenological potentials. It is suggested that
one can relate this deepening to the super-deformed state of the
compound nucleus, $^{24}$Mg.
\end{abstract}
 \maketitle
\section{Introduction}
In this paper we investigate the coupling potential that gives
rise to inelastic scattering by studying the total elastic
scattering potential that arises in the presence of inelastic
scattering. We do this by determining, by means of $S_l
\rightarrow V$ inversion, the total elastic scattering potential
corresponding to the elastic scattering $S$-matrix. This allows us
to identify the dynamical polarization potential, DPP, and it is
the properties of this which we can then relate to the
characteristics of the coupling potential.

Boztosun and Rae \cite{boztosun1} have recently made a detailed
analysis of the $^{12}$C+$^{12}$C system over a wide energy range
and have shown that the standard (conventional) coupled-channels
calculations are unable to explain the elastic and the inelastic
scattering data simultaneously and that the magnitude of the
mutual-2$^{+}$ cross-section is under-predicted by a large factor
in the $^{12}$C+$^{12}$C system. This has been a major problem for
the last couple of decades, but a new coupling potential proposed
by these authors solves many of these problems. This new coupling
potential also successfully explains elastic and inelastic
scattering of the $^{12}$C+$^{24}$Mg and $^{16}$O+$^{28}$Si
systems over a very wide range of energies
\cite{boztosun2,boztosun3,boztosun4}.

Substantial information about the effect of this new coupling
potential can be obtained from the inversion process since the new
and the standard coupled-channels calculations lead to different
elastic scattering $S$-matrices, $S_l$. This change in $S_l$ can
be represented, by inverting $S_l$ for a range of $l$ values, as a
change in the effective local elastic scattering potential, $V$.

All the required experimental data (elastic, single-$2^{+}$ and
mutual-$2^{+}$ excitations) are available at two energies: $E_{\rm
Lab}$=93.8 MeV and $E_{\rm Lab}$=126.7 MeV. These are studied with
both the conventional and the new inelastic coupling potentials.

In the next section, we review the inversion method. Sections
\ref{conexp1} and \ref{newexp1} show the results of the
conventional and new coupled-channels calculations and we present
our general conclusions in section \ref{conc}.

\section{The Method}
We apply the iterative perturbative (IP) method for $S_l$ to the
$V(r)$ inversion as implemented in the code IMAGO \cite{Cooper}.
This method has been described in detail in the references
\cite{Ait 93a,Ait 93b,Coo 90,Coo 92,Coo 96,Mac 80,Mac 82} and
relevant points and required definitions are simply noted below.
The algorithm iteratively corrects a `starting reference
potential' (SRP), here initially taken to be the square of the
Woods-Saxon potential from reference \cite{boztosun1}. The quality
of the inversion is quantified by the `phase shift difference',
$\sigma$. This is defined in terms of the target S-matrix
$S^{T}_l$ for which the potential is sought, and the $S^{I}_l$ as
calculated with the potential found by the inversion, as follows:
\begin{equation}
\sigma^{2}=\sum_{l}|S^{T}_l-S^{I}_l|^{2} \label{sig}
\end{equation}

The minimization of $\sigma$ usually requires several sequences of
inversion iterations in heavy-ion cases as presented here.

\section{Conventional Coupling Potential}
\label{conexp1}

We invert the elastic scattering $S$-matrix elements, calculated
with the standard coupling potential, at a laboratory energy of
126.7 MeV, as follows:

$(i)$ An elastic scattering calculation is carried out without any
coupling to excited states of $^{12}$C in order to verify that the
uncoupled $S$-matrix, when inverted, gives precisely what we
understand to be the `bare' potential. This also verifies that our
the coupled channel and inversion programs are numerically
consistent.  The low order moments (volume integral and RMS
radius) of the potential obtained in this way are shown in
Table~\ref{conray1}, labelled `BARE'.

$(ii)$ The first excited state (single-2$^{+}$) of the $^{12}$C is
included in a coupled channel calculation in order to observe the
effect of the inclusion of this state on the real and imaginary
components of the potential. This $S$-matrix is inverted and the
potential obtained in this way is shown in Table \ref{conray1},
labelled `SINGLE'.

$(iii)$ Mutual excitation, as well as single-2$^+$ state
excitation, is included. Initially, inversion fails to give a
non-oscillatory potential. Therefore, the `SINGLE' inverted
potential is employed as the SRP for inverting the `MUTUAL'
elastic S-matrix, and this does lead to a reasonably smooth
potential. Characteristics of the potential obtained in this way
are also shown in Table \ref{conray1}. In each case, two inverted
potentials are presented since oscillatory features tend to appear
in the potentials. These potentials are, however, much less
oscillatory than the `mutual' potential found when the bare
potential is employed as SRP. In each case, the first listed
potential is smooth but gives a less than perfect fit to $S_l$
while the second fits $S_l$ almost perfectly, albeit with some
oscillatory features.

The oscillatory features probably represent $l$-dependence in the
`true' potential, but are hard to interpret and make the
calculation of $J$ and $<r^2>^{1/2}$ problematic. It is known that
$S_l$, calculated from smooth but $l$-dependent potentials, can be
well represented by $l$-independent potentials with oscillatory
features. It is also expected from the Feshbach formalism that the
DPP will be $l$-dependent to some degree. The oscillatory features
are clearly a subject for further study, and here we simply
present two alternative fits: `smooth' and `lower $\sigma^2$',
where low $\sigma^2$ is defined in Eqn.~\ref{sig}.

Since the calculation is for identical bosons, we have $S_l$ for
even $l$ only. The inverted potentials are thus less
well-determined because there is less information to define the
potential. It may well be reasonable in future studies of this
kind to retain odd $l$ in order to either eliminate spurious
oscillatory features or to confirm that they should be present.

The nature of the DPP induced by single-channel coupling emerges
from a comparison of the characteristics of the `bare' potential,
first line in Table~\ref{conray1}, with those with single-channel
coupling, second and third lines. Single-channel coupling in this
model induces a DPP with an attractive real component with $\Delta
J_{\rm R} \sim 15$ MeV fm$^3$ and an absorptive imaginary
component with $\Delta J_{\rm I} \sim 20$ MeV fm$^3$. The change
in $<r^2>^{1/2}_{\rm R}$ is small but negative, in contrast to the
change associated with the `new' coupling, where, as we shall see
in Table \ref{newray1}, it is positive, at least for smooth
potentials.

\begin{figure}
\epsfxsize=10.0cm \centering \vskip-1cm \epsfbox{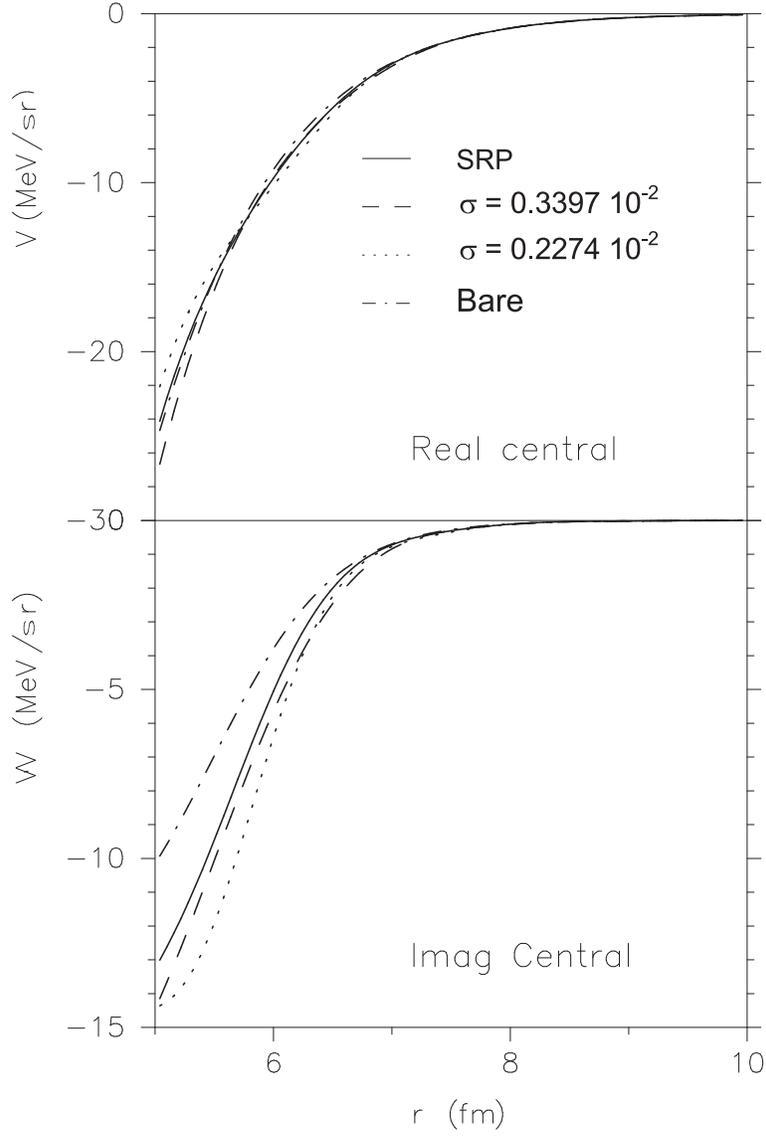}
\vskip-0cm \caption{The comparison of the potentials in the
surface region found by inverting the S-matrix of the standard
coupled-channels calculations at $E_{\rm Lab}$=126.7 MeV. The
long-dashed and the dotted lines correspond to the bottom two
lines of Table \ref{conray1} with the respective values of
$\sigma^{2}$. The dot dashed lines denote the bare potential and
finally, the solid line is the Starting Reference Potential (SRP)
for the mutual case inversion, a smooth solution to the single
case inversion.} \label{conmac1}
\end{figure}

The inclusion of  mutual excitation has a significant effect on
the potentials, as the last two lines in Table~\ref{conray1}
reveal.  For the surface region, the comparison of the new SRP and
the total inverted potentials obtained for the mutual case are
shown in Fig. \ref{conmac1} for different values of $\sigma^{2}$.
The total effect of the coupling when mutual excitation is
included is to induce a DPP with an attractive real component with
$\Delta J_{\rm R} \sim 35$ MeV fm$^3$ and an absorptive component
with $\Delta J_{\rm I} \sim 35$ MeV fm$^3$. For the real potential
in particular, the mutual coupling almost doubles the effect.

Two important outcomes have emerged at this stage: (i) the
simultaneous mutual excitation of the two nuclei in
coupled-channels calculations contributes in a substantial way to
the inter-nuclear interaction, and, (ii) it is confirmed, as is
clearly seen from Table \ref{conray1}, that inelastic coupling has
a substantial contribution to the real as well as to the imaginary
potential. Although this was pointed out long ago~\cite{rsm}, one
still often encounters the statement that inelastic coupling just
gives rise to additional absorption.

\section{The DPP with the New Coupling Potential}\label{newexp1}

\begin{figure}
\epsfxsize=12.0cm \centering \vskip-0cm \hskip-0cm
\epsfbox{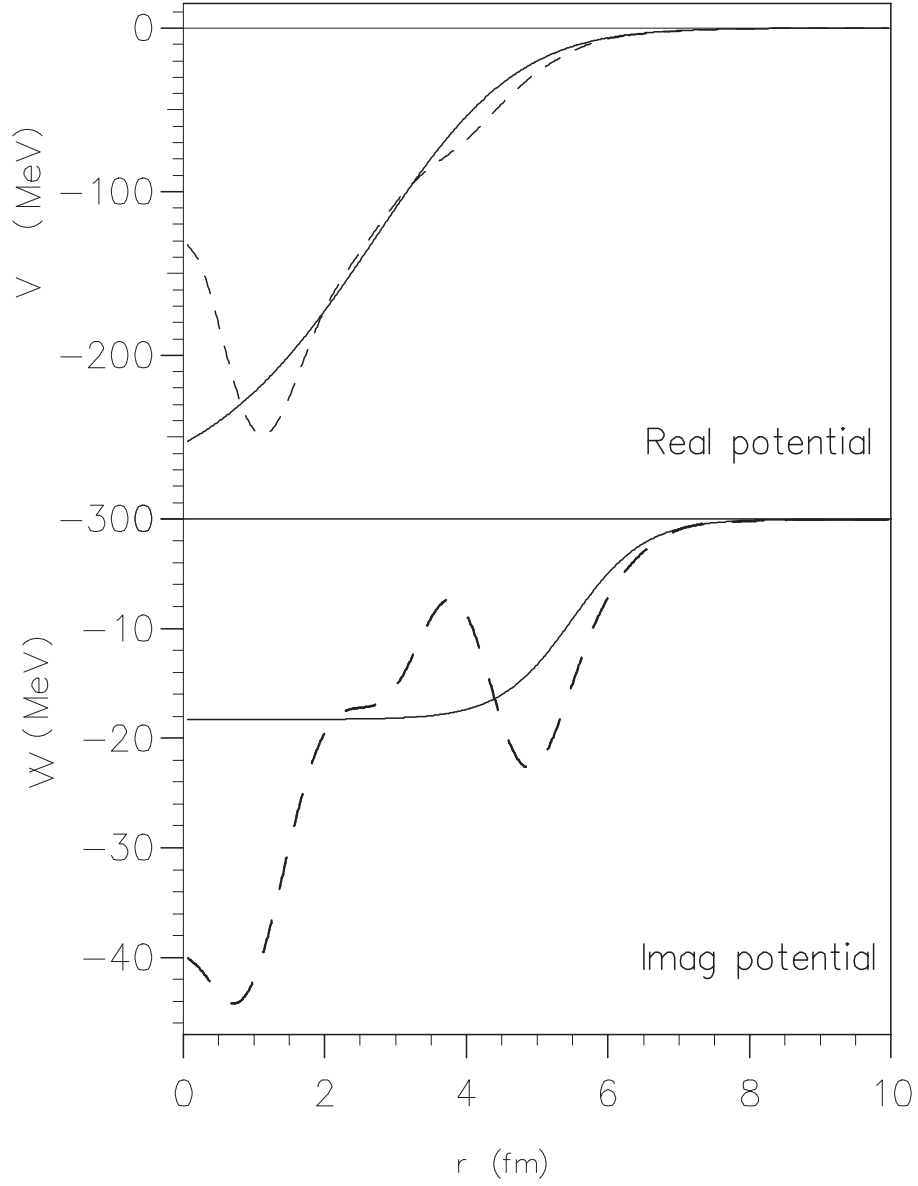} \vskip-0cm \caption{Comparison of the
total inverted potential (long dashed lines) and the bare
potential (solid lines) found by inverting the S-matrix of the new
coupled-channels calculations at $E_{\rm Lab}$=126.7 MeV. The
numerical values are given in Table \ref{newray1}.}
\label{newmac1}
\end{figure}

We now present the results of  inverting the elastic scattering
S-matrix arising from CC calculations involving the new coupling
potential. The numerical characteristics of the potentials are
shown in Table \ref{newray1} for $E_{\rm Lab}$=126.7 MeV and in
Table \ref{newtest2} for $E_{\rm Lab}$=93.8 MeV. The bare
potential and the total inverted potential are compared in Fig.
\ref{newmac1}. The new coupling potential leads to a deepening in
the surface region which can be compared with the effect of the
conventional coupling inferred from  Fig. \ref{conmac1}. This
deepening is quantified in Table \ref{newray1} where we find that
$\Delta J_{\rm R} \sim 35$ MeV fm$^3$ and $\Delta J_{\rm I} \sim
15$ MeV fm$^3$.

In contrast to the case with the conventional coupling, the added
attraction is much greater than the added absorption. It is well
known that light heavy-ion reactions are extremely sensitive to
the shape of the potential in the surface region. This deepening
in the surface region certainly has substantial effects on
scattering since it has solved many of the underlying problems of
the $^{12}$C+$^{12}$C reaction. Just why the added attraction is
much greater than the added absorption with the new coupling
interaction, and not with the standard coupling interaction, is
among the many properties of the DPP which are at present not well
understood. It presents a challenge for future studies.

Nevertheless, it could be argued that this deepening may be
interpreted in terms of the strongly deformed structure of the
target nucleus and, as Boztosun and Rae \cite{boztosun1} argued,
it may be related to the super-deformed state of the $^{24}$Mg by
considering its two $^{12}$C cluster structure.

The minimum in the surface region and the strongly deformed
structure of the target nucleus appear to be related. When one
looks at the potential of the $^{16}$O+$^{28}$Si elastic
scattering calculations of Kobos and Satchler \cite{kobos1} and
Kobos, Satchler and Mackintosh \cite{kobos2}, a minimum in the
surface region is observed. In those calculations, a deep
double-folding real potential has been used with two small {\it ad
hoc} potentials introduced artificially and they create the
minimum in the surface region. Neither Kobos and Satchler nor
others were able to fit the data without two small {\it ad hoc}
potentials.

On the other hand, Boztosun and Rae \cite{boztosun2} have also
analyzed this reaction from 29 MeV to 142.5 MeV with the new
coupling potential and successfully removed these two small {\it
ad hoc} potentials by including the single-2$^{+}$ and
single-4$^{+}$ excited states of the target nucleus, $^{28}$Si.
Kobos and Satchler interpreted their effects in terms of the
`barrier/internal' wave decomposition. However, the present
inversion results imply that the minima in the phenomenological
potentials represent inelastic coupling effects. Therefore, as we
pointed out, this phenomenon is likely to be related to the
deformation of the target nucleus, $^{28}$Si.

The whole procedure was also applied at 93.8 MeV in order to
verify the effect of the new coupling interaction on the elastic
scattering potential. Inversion proved to be less straightforward
at the lower energy, and problems are encountered in getting a
smooth potential with the `mutual' case. This is partly due to the
smaller number of partial waves contributing to the inversion and
also, possibly, due to an increased $l$-dependence at the lower
energy. Once more, a number of solutions are presented, with those
having lower $\sigma^2$ generally corresponding to more
oscillatory potentials.

The numerical values of the bare and the total inverted potentials
are shown in Table \ref{newtest2}. The first and third solutions
are much smoother than the second, which however has a notably
lower $\sigma^2$.

For the smooth potentials we have $\Delta J_{\rm R} \sim 38$ MeV
fm$^3$ and $\Delta J_{\rm I} \sim 28$ MeV fm$^3$. Thus, as at
126.7 MeV, the effect on the real part is greater than that on the
imaginary part in absolute though not in relative terms. The
different radial form of the DPP results in a tendency for
$<r^2>^{1/2}$ to increase, at least for the smooth solutions. This
is a departure from what happens with `conventional' coupling and
deserves investigation.

\section{Summary and discussion}
\label{conc}

The considerable success of the new coupling potential of Boztosun
and Rae \cite{boztosun1} has subjected the standard
coupled-channels procedure to scrutiny. Studies using this new
coupling potential may lead to new insights into the formalism and
a new interpretation of a class of direct reactions. Here we have
investigated the effect of this new coupling potential upon the
effective elastic scattering potential by inverting the elastic
$S$-matrix derived from coupled channel calculations. We have
observed that the inclusion of the excited states of the target
nucleus has an important effect on the real as well as the
imaginary potential.

When the standard coupling interaction, the added attraction is
almost the same as the added absorption. However, with the new
coupling interaction, the added attraction is much greater than
the added absorption. This deep attraction creates a deepening in
the surface region for the total inverted potential as shown in
Fig. \ref{newmac1}, and this solves many of the underlying
problems of the $^{12}$C+$^{12}$C, $^{12}$C+$^{24}$Mg and
$^{16}$O+$^{28}$Si systems over a very wide energy range
\cite{boztosun2,boztosun3,boztosun4,boztosun5}. With the standard
coupling potential, $<r^2>^{1/2}_{\rm R, I}$, $R_{\rm rms}$, tends
to decrease, but with the new coupling it tends to increase.
Understanding why $R_{\rm rms}$ increases with the new, but not
with the standard, coupling potential is a challenge for further
studies. For unstable nuclei, elastic scattering is a source of
nuclear size information, so an understanding of how inelastic
processes modify $<r^2>^{1/2}$ is important.

\section{Acknowledgments}
The authors wish to thank Drs W.D.M. Rae, Y. Nedjadi, S.
Ait-Tahar, B. Buck, A.M. Merchant and Professor B.R. Fulton for
valuable discussions and encouragements.

\newpage
\begin{table}
\begin{center}
\begin{tabular}{lccccc}\hline
Case & $\sigma^2$ & $J_{\rm R}$ & $<r^2>^{1/2}_{\rm R}$ & $J_{\rm
I}$ &
$<r^2>^{1/2}_{\rm I}$ \\
 \hline
BARE & ---- & 350.04 & 4.114 & 73.301 & 4.659 \\
SINGLE & $ 1.98\times 10^{-3}$ & 370.55 & 4.084 & 92.67 & 4.702 \\
SINGLE & $ 1.60\times 10^{-3}$ & 362.43 & 4.056 & 90.84 & 4.770 \\
MUTUAL  & $ 3.40\times 10^{-3}$ & 389.54 & 4.069 & 103.90 & 4.686 \\
MUTUAL  & $ 2.23\times 10^{-3}$ & 382.96 & 4.041 & 113.51 & 4.666 \\
\hline
\end{tabular}
\end{center}
\caption{Standard coupled-channels calculations: Numerical values
of the bare and total inverted potentials (Bare+DPP) obtained by
inverting the S-matrix at $E_{\rm Lab}$=126.7 MeV for the single
and the mutual cases, where ${\rm R}$ and ${\rm I}$ denote the
real and the imaginary parts of the potentials and of the radii
respectively.} \label{conray1}
\end{table}
\begin{table}

\begin{center}
\begin{tabular}{lccccc}\hline
Case & $\sigma^2$ & $J_{\rm R}$ & $<r^2>^{1/2}_{\rm R}$ & $J_{\rm
I}$ &
$<r^2>^{1/2}_{\rm I}$\\
\hline
BARE & ---- & 314.19 & 3.814 & 95.815 & 4.659 \\
MUTUAL  & $ 3.67\times 10^{-3}$ & 351.40 & 3.905 & 110.56 & 4.865 \\
\hline
\end{tabular}
\end{center}
\caption{New coupled-channels calculations: Numerical values of
the bare and total inverted potentials (Bare+DPP) obtained by
inverting the S-matrix at $E_{\rm Lab}$=126.7 MeV for the mutual
case, where ${\rm R}$ and ${\rm I}$ denote the real and the
imaginary parts of the potentials and of the radii respectively.}
\label{newray1}
\end{table}
\begin{table}
\begin{center}
\begin{tabular}{lccccc}\hline
Case &$\sigma^2$ & $J_{\rm R}$ & $<r^2>^{1/2}_{\rm R}$ &$J_{\rm
I}$ &
$<r^2>^{1/2}_{\rm I}$\\
\hline
BARE & ---- & 319.34 & 3.897 & 94.929 & 4.757 \\
MUTUAL  & $ 1.50\times 10^{-3}$& 356.73 & 3.919 & 123.53 & 5.099 \\
MUTUAL  & $ 1.14\times 10^{-3}$& 337.43 & 3.781 & 113.72 & 5.227 \\
MUTUAL  & $ 1.44\times 10^{-3}$& 357.88 & 3.927 & 123.84 & 5.099\\
\hline
\end{tabular}
\end{center}
\caption{New coupled-channels calculations: Numerical values of
the bare and  total inverted potentials (Bare+DPP) obtained by
inverting the S-matrix at $E_{\rm Lab}$=93.8 MeV for the mutual
case for different values of $\sigma^{2}$. ${\rm R}$ and ${\rm I}$
denote the real and the imaginary parts of the potentials and of
the radii respectively.} \label{newtest2}
\end{table}

\begin{figure}
\epsfxsize=10.0cm \centering \epsfbox{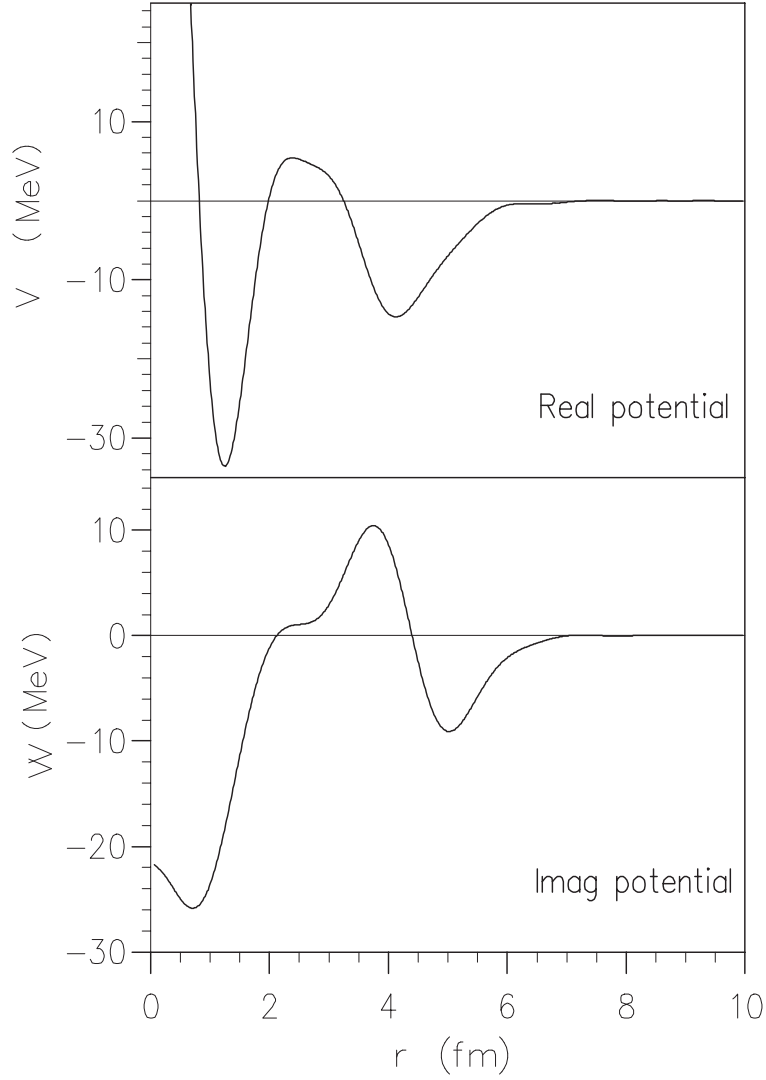} \vskip-0cm
\caption{DPP at $E_{\rm Lab}$=126.7 MeV found by subtracting the
bare potential (solid lines in Fig. \ref{newmac1}) from the total
inverted potential (long dashed lines in Fig. \ref{newmac1}). The
numerical values are given in Table \ref{newray1}.}
\label{newmac2}
\end{figure}

\end{document}